\documentclass[reprint,amsmath,amssymb,onecolumn,aps]{revtex4-2}
\usepackage{graphicx}
\usepackage{xcolor}
\usepackage{bm}
\usepackage{bbm}
\usepackage{mathrsfs}
\usepackage{slashed}
\usepackage{makecell}
\usepackage{multirow}
\allowdisplaybreaks[4]
\usepackage[colorlinks,citecolor=blue,urlcolor=blue,linkcolor=blue]{hyperref}

\begin{document}
	
	\preprint{APS/123-QED}
	
	\title{Semileptonic decay of double strangeness heavy flavor baryons}
	
	\author{Hui-Hui Duan$^1$}
	\email{duanhuihui@htu.edu.cn}
     	\author{Yong-Lu Liu$^2$}
	\email{yongluliu@nudt.edu.cn}
	\author{Qin Chang$^1$}
        \email{qinchang@htu.edu.cn}
	\author{Ming-Qiu Huang$^2$}
	\email{mqhuang@nudt.edu.cn}
	\affiliation{$^1$School of Physics, Henan Normal University, Xinxiang 453007, Henan, People's Republic of China}
	\affiliation{$^2$Department of Physics, National University of Defense Technology, Changsha 410073, Hunan, People's Republic of China}

    \begin{abstract}
   This paper investigates the double strangeness heavy flavor baryons $\Omega_c^0$ and $\Omega_b^-$, which contain two strange quarks. Using QCD light-cone sum rules (LCSRs), we calculate the form factors for the Cabibbo-suppressed processes $\Omega_c^0\to\Xi^-$ and $\Omega_b^-\to\Xi^0$, corresponding to the heavy-quark transitions $c\to d$ and $b\to u$, respectively. Combining these with the helicity amplitude formalism for semileptonic decay differential widths, we computed the branching fractions of their corresponding semileptonic decay processes. Our analysis reveals significant discrepancies between two versions of QCD LCSRs: one using the light-cone distribution amplitudes (LCDAs) of the final-state $\Xi$ baryon and the other using the LCDAs of the initial-state double strangeness heavy flavor baryons. The results obtained with the $\Xi$ baryon's LCDAs show excellent agreement with other theoretical calculations. However, when using the LCDAs of the double strangeness heavy flavor baryons, the results differ by orders of magnitude, warranting further investigation.
    \end{abstract}
	
    \maketitle

	\section{Introduction} \label{sec:I}  
	 
	The semileptonic decays of heavy flavor baryons serve as important channels for investigating their weak decay properties. In particular, the clean final states of semileptonic decays provide an excellent content for calculating weak decay form factors of heavy flavor baryons and testing various phenomenological models. Moreover, by combining experimental measurements of semileptonic branching ratios of charm and bottom hadrons, the Cabibbo-suppressed transitions $b\to u$ and $c\to d$ processes can be used to constrain the Cabibbo-Kobayashi-Maskawa (CKM) matrix elements $|V_{ub}|$ and $|V_{cd}|$. Currently, theoretical and experimental studies of these weak decays primarily focus on semileptonic and purely leptonic channels of $D$ and $B$ mesons \cite{Khodjamirian:2017fxg,Feldmann:2015xsa,Ball:2004ye,Ball:2004rg,ARGUS:1990amj}. However, investigations in the realm of heavy flavor baryons are equally important, both theoretically and experimentally.
	
	In the field of heavy flavor baryons, weak decay processes of states such as $\Lambda_Q$ and $\Xi_Q$ have been extensively studied. However, investigations of heavy flavor baryons containing two strange quarks remain relatively scarce. The weak decay parameters of double strangeness heavy flavor baryons are important for understanding the properties and structure of heavier baryonic states. In this work, we focus on the semileptonic decays of double strangeness heavy flavor baryons. Using the QCD light-cone sum rule (LCSR) approach, we theoretically investigate the weak decay form factors for the Cabibbo-suppressed transitions $\Omega_b^-\to\Xi^0$ and $\Omega_c^0\to\Xi^-$, as well as the corresponding branching fractions of the semileptonic decays $\Omega_b^-\to\Xi^0\ell^-\nu_\ell$ and $\Omega_c^0\to\Xi^-\ell^+\bar{\nu}_\ell$.
	
	Recent studies on weak decay processes of double strangeness heavy flavor baryons have seen significant experimental progress. The Belle Collaboration has reported evidence for the single Cabibbo-suppressed weak decay channel $\Omega_c^0\to\Xi^-\pi^+$ and measured its branching fraction \cite{Belle:2022yaq}. Additionally, they have also tested lepton universality in the $\Omega_c^0\to\Omega^-\ell^+\nu_\ell$  decay channel \cite{Belle:2021dgc}.

       Theoretically, investigations of $\Omega_Q$ weak decays have been conducted using various approaches, including heavy quark effective theory \cite{Du:2011nj,Singleton:1990ye,Cheng:1995fe}, the light-front quark model \cite{Zhao:2018zcb,Hsiao:2020gtc,Hsiao:2021mlp,Wang:2024mjw}, covariant confined quark model \cite{Gutsche:2018utw}, constituent quark model \cite{Pervin:2006ie},QCD sum rules \cite{Shi:2025ocl,Neishabouri:2024gbc}, QCD light-cone sum rules \cite{Aliev:2022gxi,Duan:2020xcc,Aliev:2018uby,Mannel:2011xg}, topological SU(3) \cite{Wang:2023uea}, etc.

       In this work, we employ the QCD LCSR approach to investigate the functional dependence and numerical behavior of the $\Omega_Q$ to $\Xi$ transition form factors near zero momentum transfer. By incorporating extrapolation formulas that describe the momentum-transfer dependence of these form factors, we further calculate the branching fractions for the semileptonic decay processes $\Omega_Q\to\Xi\ell\nu_\ell$. Our theoretical predictions for these form factors and branching ratios will provide valuable references for future experimental measurements of these decay channels.
	
	In the following context, Sec. \ref{secII} will present the theoretical formalism for the $\Omega_Q\to\Xi$ transition form factors derived using both the final-state $\Xi$ baryon and initial-state $\Omega_Q$ baryon light-cone distribution amplitudes (LCDAs), along with the helicity amplitude formalism for calculating differential decay widths of semileptonic decays. Section \ref{sec:III} provides numerical results showing the momentum-transfer dependence of the form factors, together with the calculated decay widths and absolute branching fractions for the semileptonic transitions. The paper concludes with a comprehensive summary and discussion in the final section.
	
	 \section{Form factors of $\Omega_Q \to \Xi$ \label{secII}}
    
     In this section, we calculate the form factors of weak decay processes using the theoretical framework of QCD LCSRs. In the computation of the form factors, we separately consider the theoretical results obtained when employing the LCDAs of the initial- and final-state baryons. The LCDAs for the $\Xi$ baryon can be found in Ref. \cite{Liu:2008yg}, while those for the $\Omega_Q$ baryon  are provided in Ref. \cite{Ali:2012pn}. Unlike our previous works \cite{Duan:2022yia, Duan:2022uzm}, where we calculated the form factors using either the final- or initial-state baryon LCDAs and considered the contributions from negative parity baryons, in this study, we follow the approach of Refs. \cite{Huang:2004vf,Wang:2009hra} and discuss the transition form factors by incorporating both the final- and initial-state baryon LCDAs.
    
\subsection{Theoretical frame of light-cone sum rules}

    According to quantum chromodynamics and the Standard Model of particle physics, the form factors of baryonic weak decays can be defined by parametrizing the transition matrix elements of the weak decay $V-A$ current in terms of three vector form factors and three axial-vector form factors, expressed as
  \begin{align}
    	\langle \Omega_Q&(P_{\Omega_Q})|j_1^\nu|\Xi(p)\rangle \nonumber \\
    	&=\bar{u}_{\Omega_Q}(p)\left[f_1(q^2)\gamma^\nu+ i \frac{f_2(q^2)}{M_{\Omega_Q}} \sigma^{\nu\mu}q_\mu + \frac{f_3(q^2)}{M_{\Omega_Q}}q^\nu\right. \nonumber \\&\left. -\left(g_1(q^2)\gamma^\nu+i \frac{g_2(q^2)}{M_{\Omega_Q}} \sigma^{\nu\mu} q_\mu +\frac{g_3(q^2)}{M_{\Omega_Q}}q^\nu\right)\gamma_5\right]u_{\Xi}(p), \label{ff1}
  \end{align}
  or
    \begin{align}
    	\langle \Xi&(p)|j_2^\nu|\Omega_Q(P_{\Omega_Q})\rangle \nonumber \\
    	&=\bar{u}_{\Xi}(p)\left[f_1(q^2)\gamma^\nu+ i \frac{f_2(q^2)}{M_{\Omega_Q}} \sigma^{\nu\mu}q_\mu + \frac{f_3(q^2)}{M_{\Omega_Q}}q^\nu\right. \nonumber \\&\left. -\left(g_1(q^2)\gamma^\nu+i \frac{g_2(q^2)}{M_{\Omega_Q}} \sigma^{\nu\mu} q_\mu +\frac{g_3(q^2)}{M_{\Omega_Q}}q^\nu\right)\gamma_5\right]u_{\Omega_Q}(P_{\Omega_Q}). \label{ff2}
	\end{align}
Where, $f_i$  and $g_i$ $(i=1,2,3)$ denote the weak decay form factors, $M_{\Omega_Q}$  is the mass of the double strangeness heavy baryon $\Omega_Q$, $q$ is the momentum transfer in the $\Omega_Q \to \Xi$ transition, $p$ is the momentum of the $\Xi$ baryon, and $P_{\Omega_Q}=p-q$ is the momentum of the $\Omega_Q$ baryon. We assume that the double strangeness heavy baryon $\Omega_Q$ is on shell.
  
To extract information about hadronic weak decay form factors within the LCSR approach, one analyzes a matrix element in which one hadron is represented by an interpolating field with appropriate quantum numbers, while the other is described by its explicit state vector \cite{Braun:2005be}. When constructing the sum rule, the distribution amplitudes (DAs) of the latter hadron are required. In the decay process of $\Omega_Q \to\Xi$, there are two hadrons involved: the initial-state $\Omega_Q$ and the final-state $\Xi$. The concept of hadron distribution amplitudes generally refers to hadron-to-vacuum matrix elements of nonlocal operators composed of quark and gluon fields at lightlike separations \cite{Braun:2000kw}. Based on the aforementioned construction approach, when performing sum rule analysis using the DAs of the final-state $\Xi$, the correlation function takes the following form:
       \begin{gather}
    	T^\nu (P,q)=i\int d^4x e^{iq \cdot x}\langle 0|T\{j_{\Omega_Q}(x)j_1^\nu (0)\}|\Xi(p)\rangle ; \label{eq3}
       \end{gather}
this differs from the conventional representation of the hadronic weak decay matrix element. Typically, the initial hadronic state is placed in the ket. However, in order to utilize the form of the $\Xi$ baryon DAs provided in Ref. \cite{Liu:2008yg} in subsequent LCSR calculations, we place the final-state $\Xi$ baryon in the ket. Consequently, we apply a complex conjugate transformation to the transition matrix element of $\Omega_Q \to \Xi$, which leads to the form of the correlation function given in Eq. (\ref{eq3}). In the subsequent calculations, the correlation function in Eq. (\ref{eq3}) corresponds to the form factors defined in Eq. (\ref{ff1}). Furthermore, since the form factors defined in this work are all real, the complex conjugate transformation does not affect the form factors.
When performing sum rule analysis using the DAs of the initial-state $\Omega_Q$, the correlation function is constructed in the following form:
        \begin{gather}
    	T^\nu (P,q)=i\int d^4x e^{ip \cdot x}\langle 0|T\{j_{\Xi}(x)j_2^\nu (0)\}|\Omega_Q(P_{\Omega_Q})\rangle , \label{eq4}
       \end{gather}
and this correlation function corresponds to the form factors defined in Eq. (\ref{ff2}). $j_\Xi(x)$ and $j_{\Omega_Q}(x)$ in the correlation functions Eq. (\ref{eq3}) and (\ref{eq4}) are the interpolating currents corresponding to the $\Xi$ and $\Omega_Q$ baryons with the appropriate quantum numbers, and $j_{1(2)}^\nu$ represents the $V$-$A$ current of weak decays. In this work, we adopt the Chernyak-Zhitnitsky type interpolating current for the baryons,
	   \begin{align}
    	j_{\Xi}(x)=&\epsilon_{ijk}(s^{iT}(x)C\slashed{z} s^j(x))\gamma_5 \slashed{z} q^k(x),  \nonumber \\
    	j_{\Omega_Q}(x)=&\epsilon_{ijk}\left(s^{iT}(x)C\slashed{z}s^j(x)\right)\gamma_5\slashed{z} Q^k(x).
    	   \end{align} 
	Here, we introduce a light-cone vector $z_\nu$ satisfying $z_\nu z^\nu=z^2=0$. The $V$-$A$ current of weak decay is given by:
	   \begin{align}
          j^\nu_1(x)=\bar{Q}(x)\gamma_\nu(1-\gamma_5)q(x), \nonumber \\
          j^\nu_2(x)=\bar{q}(x)\gamma_\nu(1-\gamma_5)Q(x).
	    \end{align}
	
	  The matrix element of the baryon transition to the vacuum state can be defined via a decay constant:
	    \begin{align}
	\langle 0|j_{\Omega_Q}(x) |\Omega_Q(P_{\Omega_Q}) \rangle =&  f_{\Omega_Q} z\cdot P_{\Omega_Q}\slashed{z}u_{\Omega_Q} (P_{\Omega_Q}), \nonumber \\
	\langle 0|j_\Xi(x)|\Xi(p)\rangle=&f_\Xi z\cdot p\slashed{z}u_\Xi(p),
	    \end{align}
where $f_{\Omega_Q}$ and $f_\Xi$ are the decay constants of the double strangeness heavy flavor baryon $\Omega_Q$  and the light flavor baryon $\Xi$, respectively. Based on these definitions, we compute the correlation function on both the hadronic and QCD levels and match the results to extract the weak decay form factors.

First, on the hadronic level, we insert a complete set of intermediate states with the same quantum numbers as $\Xi$ and $\Omega_Q$  into the correlation functions. In the framework of QCD LCSRs, we perform the operator product expansion of the correlation function near the light cone (i.e., $z^2=0$ or $x^2=0$). The calculation of weak decay form factors using LCSR is only valid when the momentum transfer $q^2$ lies close to $0~\rm{GeV}^2$. In this regime, the projected momentum transfer along the light cone constitutes a subdominant quantity compared to the momenta of the initial- and final-state hadrons. Therefore, by multiplying both sides of the hadronic and QCD representations by the light-cone vector $z_\nu$ and projecting the momenta onto the light cone, we can neglect the subdominant small $q\cdot z$ term associated with lepton momenta in semileptonic decays. The form factors $f_3$ and $g_3$ contribute to the absolute branching fraction of semileptonic decays only when accounting for mass differences among the final-state leptons.  Assuming negligible lepton mass effects, the form factors $f_3(q^2)$ and $g_3(q^2)$ do not contribute to the sum rules. Under these considerations, the hadronic representations of the correlation functions are given by
 \begin{align}
 z_\nu T^\nu(p,q)=&\frac{f_{\Omega_Q}}{M_{\Omega_Q}^2-P_{\Omega_Q}^2}2(z\cdot p)^2\left\{f_1\slashed{z}-\frac{f_2}{M_{\Omega_Q}}\slashed{z}\slashed{q}
 -g_1\slashed{z}\gamma_5+\frac{g_2}{M_{\Omega_Q}}\slashed{z}\slashed{q}\gamma_5\right\}u(p)+\cdots \label{cor1}
 \end{align}
 or
 \begin{align}
 z_\nu T^\nu(p,q)=&\frac{f_\Xi}{M_\Xi^2-p^2}2(p\cdot z)^2\left\{f_1\slashed{z}-\frac{f_2}{M_{\Omega_Q}}\slashed{z}\slashed{q}
 -g_1\slashed{z}\gamma_5+\frac{g_2}{M_{\Omega_Q}}\slashed{z}\slashed{q}\gamma_5\right\}u(P_{\Omega_Q})+\cdots, \label{cor2}
 \end{align}
where the ellipses denote contributions from excited states and continuum spectra. Here, we have used the spin summation relations $\sum_s u_\Xi(p)\bar{u}_\Xi(p)=(\slashed{p}+M_\Xi)$ and $\sum_s u_{\Omega_Q}(P_{\Omega_Q})\bar{u}_{\Omega_Q}(P_{\Omega_Q})=(\slashed{P}_{\Omega_Q}+M_{\Omega_Q})$.

Next, we derive the QCD-level expressions for the correlation functions. By substituting the baryonic interpolating currents and the weak $V$-$A$ current into the correlation functions (\ref{cor1}) and (\ref{cor2}) and applying Wick’s theorem along with the free quark propagator,
     \begin{gather}
     	S^{q/Q} (x)=i\int \frac{d^4k}{(2\pi)^4}\frac{\slashed{k}+m_{q/Q}}{k^2-m_{q/Q}^2}e^{-ikx},
     \end{gather}
      we obtain the QCD representations,
\begin{align}
    z_\nu T^\nu&=\int d^4x \frac{d^4 k}{(2\pi)^4}\frac{e^{i(q+k)}}{k^2-m_Q^2}\gamma_5\slashed{z}(\slashed{k}+m_Q)\slashed{z}(1-\gamma_5) \langle 0|s(x)C\slashed{z} s(x)q(0)|\Xi(p)\rangle  \label{corQCD2}
\end{align}    
or
 \begin{align}
         z_\nu T^\nu&=\int d^4 x  \int \frac{d^4 k}{(2\pi)^4} \frac{e^{i(p-k)\cdot x}}{k^2-m_q^2} \gamma_5\slashed{z}(\slashed{k}+m_q)\slashed{z}(1-\gamma_5)  \langle 0 | s(x)C\slashed{z} s(x) Q(0) |\Omega_Q (P_{\Omega_Q})\rangle. \label{corQCD1}
        \end{align}
The matrix elements $\langle 0|s(x)C\slashed{z} s(x)q(0)|\Xi(p)\rangle$ and $\langle 0|s(x)C\slashed{z} s(x)Q(0)|\Omega_Q(P_{\Omega_Q})\rangle$ can be parametrized in terms of the LCDAs of $\Xi$ and $\Omega_Q$ baryons, respectively.

In the following, we will use these two representations to derive the theoretical expressions for the form factors of the double strangeness heavy baryon weak decays.
	
	\subsection{Form factors of $\Omega_Q\to \Xi$ within $\Xi$ LCDAs}

     When computing the form factors using the $\Xi$ baryon LCDAs in the QCD-level correlation function Eq. (\ref{corQCD1}), we adopt the parametrizations provided in Ref. \cite{Liu:2008yg}. This approach has previously yielded results in good agreement with experimental data for 
$\Xi_c$ transition to $\Xi$ semileptonic decays \cite{Duan:2022yia}. Following the methodology of Ref. \cite{Huang:2004vf}, we apply a Borel transformation to both the hadronic and QCD representations of the correlation function, introducing a Borel parameter $M_B$. By incorporating the contributions from excited states and continuum spectra via a threshold parameter $s_0$ in the dispersion relation, the theoretical expressions for the form factors are derived as    
      \begin{align}
      \frac{1}{2f_{\Omega_Q}}e^{-M_{\Omega_Q}/M_B^2}f_1(q^2)&=-\int_{\alpha_{30}}^{1}\frac{\rho_{f_{11}}^\prime(\alpha_3)}{\alpha_3}d\alpha_3e^{-s/M_B^2} \notag \\ 
               &+\frac{1}{M_B^2}\int_{\alpha_{30}}^{1}\frac{\rho_{f_{12}}^\prime(\alpha_3)}{\alpha_3^2}d\alpha_3e^{-s/M_B^2}+\frac{\rho_{f_{12}}^\prime(\alpha_{30})e^{-s_0/M_B^2}}{\alpha_{30}^2M_\Xi^2-q^2+m_Q^2} \notag  \\
               &-\frac{1}{2M_B^4}\int_{\alpha_{30}}^{1}\frac{\rho_{f_{13}}^\prime(\alpha_3)}{\alpha_3^3}e^{-s/M_B^2}d\alpha_3 -\frac{1}{2}\frac{\rho_{f_{13}}^{\prime}(\alpha_{30})e^{-s_0/M_B^2}}{\alpha_{30}M_B^2(\alpha_{30}^2M_\Xi^2-q^2+m_Q^2)} \notag \\
               &+\frac{1}{2}\frac{\alpha_{30}^2}{\alpha_{30}^2M_\Xi^2-q^2+m_Q^2}\frac{d}{dx}\frac{\rho_{f_{13}}^\prime(x)}{x(x^2M_\Xi^2-q^2+m_Q^2)}\big|_{x\to\alpha_{30}}e^{-s_0/M_B^2},
      \end{align}

      \begin{align}
      \frac{-M_{\Omega_{Q}}}{2f_{\Omega_Q}}e^{-M_{\Omega_Q}/M_B^2}f_2(q^2)&=-\int_{\alpha_{30}}^{1}\frac{\rho_{f_{21}}^\prime(\alpha_3)}{\alpha_3}d\alpha_3e^{-s/M_B^2} \notag \\ 
               &+\frac{1}{M_B^2}\int_{\alpha_{30}}^{1}\frac{\rho_{f_{22}}^\prime(\alpha_3)}{\alpha_3^2}d\alpha_3e^{-s/M_B^2}+\frac{\rho_{f_{22}}^\prime(\alpha_{30})e^{-s_0/M_B^2}}{\alpha_{30}^2M_\Xi^2-q^2+m_Q^2} \notag  \\
               &-\frac{1}{2M_B^4}\int_{\alpha_{30}}^{1}\frac{\rho_{f_{23}}^\prime(\alpha_3)}{\alpha_3^3}e^{-s/M_B^2}d\alpha_3 -\frac{1}{2}\frac{\rho_{f_{23}}^{\prime}(\alpha_{30})e^{-s_0/M_B^2}}{\alpha_{30}M_B^2(\alpha_{30}^2M_\Xi^2-q^2+m_Q^2)} \notag \\
               &+\frac{1}{2}\frac{\alpha_{30}^2}{\alpha_{30}^2M_\Xi^2-q^2+m_Q^2}\frac{d}{dx}\frac{\rho_{f_{23}}^\prime(x)}{x(x^2M_\Xi^2-q^2+m_Q^2)}\big|_{x\to\alpha_{30}}e^{-s_0/M_B^2},
      \end{align}
where
\begin{align}
\rho^\prime_{f_{11}}(\alpha_3)=&2D_0(\alpha_3)\alpha_3, \nonumber \\
\rho^\prime_{f_{12}}(\alpha_3)=&\left[D_3(\alpha_3)+D_4(\alpha_3)-2D_1(\alpha_3)\right]M_\Xi^2\alpha_3^2, \nonumber \\
\rho^\prime_{f_{13}}(\alpha_3)=&-4D_5(\alpha_3)M_\Xi^4\alpha_3^3, \nonumber \\
\rho^\prime_{f_{21}}(\alpha_3)=&0, \nonumber \\
\rho^\prime_{f_{22}}(\alpha_3)=&2D_1(\alpha_3)M_\Xi\alpha_3, \nonumber \\
\rho^\prime_{f_{23}}(\alpha_3)=&4D_5(\alpha_3)M_\Xi^3\alpha_3^2,
\end{align}
with
\begin{align}
D_0(\alpha_3)=&\int_0^{1-\alpha_3}d\alpha_1 V_1(\alpha), \nonumber \\
D_1(\alpha_3)=&\int_0^{\alpha_3}d\alpha_3^\prime \int_0^{1-\alpha_3^\prime}d\alpha_1 \left( V_1(\alpha^\prime)-V_2(\alpha^\prime) -V_3(\alpha^\prime)\right), \nonumber \\
D_2(\alpha_3)=&\int_0^{1-\alpha_3}d\alpha_1 V_3(\alpha), \nonumber \\
D_3(\alpha_3)=&\int_0^{\alpha_3}d\alpha_3^\prime \int_0^{1-\alpha_3^\prime}d\alpha_1\left(\right.-2V_1(\alpha^\prime)+V_3(\alpha^\prime) 
+V_4(\alpha^\prime)+2V_5(\alpha^\prime)\left.\right), \nonumber \\
D_4(\alpha_3)=&\int_0^{\alpha_3}d\alpha_3^\prime\int_0^{1-\alpha_3^\prime}d\alpha_1(V_4(\alpha^\prime)-V_3(\alpha^\prime)), \nonumber \\
D_5(\alpha_3)=&\int_0^{\alpha_3}d\alpha_3^\prime\int_0^{\alpha_3^\prime}d\alpha_3^{\prime\prime}\int_0^{1-\alpha_3^{\prime\prime}}d\alpha_1(-V_1(\alpha^{\prime\prime})+V_2(\alpha^{\prime\prime}) \nonumber \\ &+V_3(\alpha^{\prime\prime})+V_4(\alpha^{\prime\prime}) +V_5(\alpha^{\prime\prime})-V_6(\alpha^{\prime\prime})),
\end{align}
where $\alpha=(\alpha_1, 1-\alpha_3-\alpha_1, \alpha_3)$, $\alpha^\prime=(\alpha_1, 1-\alpha_3^\prime-\alpha_1, \alpha_3^\prime)$, and $\alpha_3^{\prime\prime}=(\alpha_1, 1-\alpha_3^{\prime\prime}-\alpha_1, \alpha_3^{\prime\prime})$. The definitions and explicit forms of $V_1(\alpha), V_2(\alpha), V_3(\alpha), V_4(\alpha), V_5(\alpha)$, and $V_6(\alpha)$ are consistent with those in Refs. \cite{Liu:2008yg, Duan:2022yia}.The axial-vector form factors $g_1(q^2)$ and $g_2(q^2)$ have the relations with $f_1(q^2)$ and $f_2(q^2)$:$g_1(q^2)=f_1(q^2)$ and $g_2(q^2)=-f_2(q^2)$.
	
	 \subsection{Form factors of $\Omega_Q\to\Xi$ within $\Omega_Q$ LCDAs }

Before calculating the form factors for the $\Omega_Q$ baryon transition to the $\Xi$ baryon using the LCDAs of the $\Omega_Q$ baryon, we first discuss the LCDAs of $\Omega_Q$. Currently, the LCDAs of double strangeness heavy flavor baryons have only been explored for the $\Omega_b$ baryon in the heavy quark limit and the rest frame of the bottom quark \cite{Ali:2012pn}. Studies on the LCDAs of charm-containing double strangeness heavy flavor baryons are relatively scarce. However, to some extent, the charm quark can also be treated as a heavy quark, and its LCDAs can be discussed analogously to those of the bottom-containing double strangeness heavy baryons in the heavy quark limit. Under this premise, the parameters and theoretical forms of the LCDAs for $\Omega_b$ and $\Omega_c$ differ only in the nonperturbative parameters—namely, the decay constants of $\Omega_c$ and $\Omega_b$.

In Ref. \cite{Ali:2012pn}, the matrix element of the $\Omega_b$ baryon decaying to the vacuum state, expressed in terms of its LCDAs, is given as
      \begin{align}
      	\langle 0 |q_1(t_1n)&\mathcal{C}\Gamma q_2(t_2n)Q_{\gamma}(0)|H_b(v)\rangle  \nonumber \\& =\frac{1}{4\sqrt{3}}{\rm{Tr}}\left[\Gamma \cdot\left(f^{(2)}_{H_b^{j=1}}\frac{i}{2}(v_+ \sigma_{\bar{n}\epsilon_\perp}\Psi^n_\perp(t_1,t_2)+e \sigma_{n\bar{n}}\Psi_{||}^{n\bar{n}}(t_1,t_2)  +\frac{1}{v_+}\sigma_{n\epsilon_\perp}\Psi^{\bar{n}}_\perp(t_1,t_2)\right)\right. \nonumber \\ &
      	+e f^{(2)}_{H_b^{j=1}}\Psi^{\mathbbm{1}}_{||}(t_1,t_2)+\frac{i}{2}f^{(1)}_{H_b^{j=1}}\gamma_5\gamma_\alpha\epsilon^{\alpha\epsilon_\perp n\bar{n}}\Psi^{n\bar{n}}_\perp(t_1,t_2) \nonumber \\ & 
      	\left.\left.+f^{(1)}_{H_b^{j=1}}\left(\frac{ev_+}{2}\slashed{\bar{n}}\Psi^n_{||}(t_1,t_2)+\slashed{\epsilon}_\perp \Psi_\perp^\mathbbm{1}(t_1,t_2)  -\frac{e}{2v_+}\slashed{n} \Psi^{\bar{n}}_{||}(t_1,t_2)\right)\right)\right]u_\gamma.
      \end{align}
      However, a more commonly used parametrization in the literature \cite{Aliev:2022gxi}, considering equal contributions from the transverse and longitudinal polarizations of the diquark system composed of two strange quarks is
      \begin{align}
      \epsilon_{ijk}\langle 0| & q^i_\alpha (t_1n)q^j_\beta(t_2n)Q^k_{\gamma}(0)| \Omega_Q\rangle \notag \\ 
     &= \frac{1}{8}v_+f_{\Omega_Q}^{(1)}\Psi^n(t_1,t_2)(\slashed{\bar{n}}\gamma_5C^T)_{\beta\alpha}u_{\Omega_Q\gamma} \notag \\
     &+\frac{1}{4}f_{\Omega_Q}^{(2)}\Psi^{\mathbbm{1}}(t_1,t_2)(\gamma_5C^T)_{\beta\alpha}u_{\Omega_Q\gamma} \notag \\  
     & -\frac{1}{8}f_{\Omega_Q}^{(2)}\Psi^{n\bar{n}}(t_1,t_2)(i\sigma_{n\bar{n}}\gamma_5C^T)_{\beta\alpha}u_{\Omega_Q\gamma} \notag \\
     &+ \frac{1}{8}\frac{1}{v_+}f_{\Omega_Q}^{(1)}\Psi^{\bar{n}}(t_1,t_2)(\slashed{n}\gamma_5C^T)_{\beta\alpha}u_{\Omega_Q\gamma}.  
      \end{align}    
     In this work, we adopt this parametrization form of the baryon decay matrix element to calculate the form factors. Here,
      \begin{gather}
      	n_\mu=\frac{x_\mu}{v\cdot x},\qquad \overline{n}_\mu=2v_\mu-\frac{x_\mu}{v\cdot x}, \qquad \overline{v}_\mu=\frac{x_\mu}{v\cdot x}-v_\mu,
      \end{gather}
    with $t_1=t_2=v\cdot x$, where $v$ is the velocity of the $\Omega_Q$ baryon, and $x$ is the position coordinate of the internal light quark. The LCDAs are expressed as
      \begin{align}
      	\tilde{\psi}_2(\omega,u)&=\omega^2 \overline{u}u\sum_{n=0}^{2}\frac{a_n}{\epsilon^4_n}\frac{C^{3/2}_n(2u-1)}{|C^{3/2}_n|^2}e^{-\omega/\epsilon_n}, \\
      	\tilde{\psi}_3^{\sigma,s}(\omega,u)&=\frac{\omega}{2}\sum_{n=0}^{2}\frac{a_n}{\epsilon_n^3}\frac{C_n^{1/2}(2u-1)}{|C_n^{1/2}|^2}e^{-\omega/\epsilon_n}, \\
      	\tilde{\psi}_4(\omega,u)&=\sum_{n=0}^{2}\frac{a_n}{\epsilon_n^2}\frac{C_n^{1/2}(2u-1)}{|C_n^{1/2}|^2}e^{-\omega/\epsilon_n}.
      \end{align}
    The wave functions $\tilde{\psi}_2, \tilde{\psi}_3^s, \tilde{\psi}_3^\sigma, \tilde{\psi}_4$ are the LCDAs of the $\Omega_Q$ baryon have the definite twists corresponding with $\Psi_{\parallel/\perp}^n, \Psi_{\parallel/\perp}^{\mathbbm{1}}, \Psi_{\parallel/\perp}^{n\bar{n}}, \Psi_{\parallel/\perp}^{\bar{n}}$, and the lower index numbers $2, 3$ and $4$ represent the twist of LCDAs. The parameters $\epsilon_i$ and$a_i$ in the distribution amplitudes are list in Table~\ref{table 1}. The Gegenbauer polynomial $C_n^\lambda$ is given by
  \begin{gather}
  	|C_n^\lambda|^2=\int_{0}^{1}du[C_n^\lambda(2u-1)]^2,
  \end{gather}
  with $|C_0^{1/2}|^2=|C_0^{3/2}|^2=1$, $|C_1^{1/2}|^2=1/3$, $|C_1^{3/2}|^2=3$, $|C_2^{1/2}|^2=1/5$, $|C_2^{3/2}|^2=6$, and $C_0^\lambda(x)=1$, $C_1^\lambda(x)=2\lambda x$, and $C_2^\lambda(x)=2\lambda(1+\lambda)x^2-\lambda$.
  
    \begin{table}[htbp]
    	\centering
    	\caption{The parameters of $\Omega_Q$ baryon distribution amplitudes to twist 4.}  \label{table 1}
    	\begin{tabular}{{ccccccc}}\hline
    		Twist&$a_0$&$a_1$&$a_2$&$\epsilon_0$(GeV)&$\epsilon_1$(GeV)&$\epsilon_2$(GeV) \\ \hline
    		2&1&$\cdots$&$\frac{8A+1}{A+1}$&$\frac{1.3A+1.3}{A+6.9}$&$\cdots$&$\frac{0.41A+0.06}{A+0.11}$ \\ 
    		$3s$&1&$\cdots$&$\frac{0.17A-0.16}{A-2}$&$\frac{0.56A-1.1}{A-3.22}$&$\cdots$&$\frac{0.44A-0.43}{A+0.27}$ \\ 
    		$3a$&$\cdots$&1&$\cdots$&$\cdots$&$\frac{0.45A-0.63}{A-1.4}$&$\cdots$ \\ 
    		4&1&$\cdots$&$\frac{-0.10A-0.01}{A+1}$&$\frac{0.62A+0.62}{A+1.62}$&$\cdots$&$\frac{0.87A+0.07}{A+2.53}$ \\ \hline
    	\end{tabular}
    \end{table}
  
  By analogy with the LCDAs of the $\Omega_b$ baryon, the LCDAs of $\Omega_c$ can be expressed in a similar form. At the leading order of heavy quark effective theory and omit the $1/m_Q$ correction in the $m_Q \to \infty$ limit, we have the relation $|\Omega_Q(P_{\Omega_Q})\rangle =\sqrt{M_{\Omega_Q}}|H_Q(v)\rangle$. With the hadronic representations and the LCDAs of the $\Omega_Q$ baryon, we can safely substitute the $|H_Q(v)\rangle$ with $|\Omega_Q(P_{\Omega_Q})\rangle$.
	
	When calculating the $\Omega_Q\to\Xi$ form factors by using the LCDAs of the $\Omega_Q$ baryon, it is still necessary to perform a Borel transformation on the correlation functions on both the hadronic and QCD levels and match them via dispersion relations. The relevant Borel transformation formula in this step is
        \begin{align}
        	\int_0^\infty d\sigma & \frac{\sigma^{n+1}\hat{\rho}^{(i)}_n (\sigma)}{[(q-\omega v)^2-m_q^2]^{n+1}}\to 
        	 \int^{\sigma_0}_0 d\sigma \frac{\rho^{(i)}_n(\sigma)e^{-s(\sigma)/M_B^2}}{(M_B^2)^n}
        	 +e^{-s_0(\sigma_0)/M_B^2}\sum_{\mathit{l}=0}^{n-1}\frac{\eta(\sigma_0)\mathcal{D}^{\mathit{l}}_\eta [\rho^{(i)}_n](\sigma_0)}{(M_B^2)^{n-\mathit{l}-1}},
        \end{align}
  where we have
  \begin{gather}
  	\hat{\rho}^{(i)}_n (\sigma)=(-1)^{n+1}n!\rho^{(i)}_n (\sigma).
  \end{gather}
  For $n=0,1,2$, we have the following formulas:
 
  $n=0$,
  \begin{align}
  \int_{0}^{\infty}d\sigma&\frac{\rho_0^{(i)}(\sigma)}{(p-\sigma M_{\Omega_Q} v)^2-m_q^2}\to 
   -\int_{0}^{\sigma_0}d\sigma\frac{1}{\overline{\sigma}}\rho_0^{(i)}e^{-s/M_B^2},
  \end{align}
  
  $n=1$,
  \begin{align}
  \int_{0}^{\infty}d\sigma & \frac{\rho_1^{(i)}(\sigma)}{[(p-\sigma M_{\Omega_Q} v)^2-m_q^2]^2}\to \\ \notag & 
  +\int_{0}^{\sigma_0}d\sigma\frac{1}{\overline{\sigma}^2}\frac{\rho_1^{(i)}(\sigma)}{M_B^2}e^{-s/M_B^2}
  +\frac{1}{\overline{\sigma}_0^2}e^{-s_0/M_B^2}\eta(\sigma_0)\rho_1^{(i)}(\sigma_0),
  \end{align}
  
  $n=2$,
  \begin{align}
  \int_{0}^{\infty}d\sigma & \frac{\rho_2^{(i)}(\sigma)}{[(p-\sigma M_{\Omega_Q} v)^2-m_q^2]^3}\to \notag \\  &-\int_{0}^{\sigma_0}d\sigma\frac{1}{2\overline{\sigma}^3}\frac{\rho_2^{(i)}(\sigma)}{M_B^4}e^{-s/M_B^2}
   -\frac{1}{2\overline{\sigma}_0^3}e^{-s_0/M_B^2}\eta(\sigma_0)\rho_2^{(i)}(\sigma_0)  \notag \\ &
  -\frac{1}{2\overline{\sigma}_0^3}e^{-s_0/M_B^2}\eta(\sigma_0)\frac{d}{d\sigma}[\eta(\sigma)\rho_2^{(i)}(\sigma)]_{\sigma=\sigma_0},
  \end{align}
where we have the parameters defined as
\begin{gather}
	\sigma=\frac{\omega}{M_{\Omega_Q}} \qquad s=\sigma M_{\Omega_Q}^2+\frac{m_q^2-\sigma q^2}{\overline{\sigma}}, \qquad \overline{\sigma}=1-\sigma,
\end{gather}
   and
   \begin{gather}
     \eta(\sigma)\equiv\frac{d\sigma}{ds}=\frac{\overline{\sigma}^2}{\overline{\sigma}^2 M_{\Omega_Q} ^2+m_q^2-q^2}.	
    \end{gather}
 The definition of $\omega$ is the energy of light diquark, and 
 \begin{align}
 	\sigma_0=&\frac{(s_0+M_{\Omega_Q}^2-q^2)}{2M_{\Omega_Q}^2} 
 	- \frac{\sqrt{(s_0+M_{\Omega_Q}^2-q^2)^2-4M_{\Omega_Q}^2(s_0-m_q^2)}}{2M_{\Omega_Q}^2}.
 \end{align}
  $s_0$ is the threshold for incorporating the contribution of the excited and continuum states.
   
   Finally, the expression for the form factors in this framework is obtained as
    \begin{align}
    	f_1(q^2)&=\left\{-\int_{0}^{1}du\int_{0}^{\sigma_0}d\sigma\frac{\rho_{11}(\sigma,u,q^2)}{1-\sigma}e^{(M_\Xi^2-s)/M^2}\right.  \notag \\&
               +\int_{0}^{1}du  \int_{0}^{\sigma_0}d\sigma\frac{\rho_{12}(\sigma,u,q^2)}{(1-\sigma)^2M_B^2}e^{(M_{\Xi}^2-s)/M_B^2} \notag \\& 
               \left.+\int_{0}^{1}du\frac{\rho_{12}(\sigma_0,u,q^2)\eta(\sigma_0,q^2)}{(1-\sigma_0)^2}e^{(M_{\Xi}^2-s)/M_B^2}\right\}\frac{f_{\Omega_Q}}{f_{\Xi}},
    \end{align}
   where the $\rho_{11}$ and $\rho_{12}$ expressions are
    \begin{align}
        \rho_{11}(\sigma,u,q^2)&=-\sigma (1-\sigma )M_{\Omega_Q}\psi^n(\omega,u),  \notag \\
        \rho_{12}(\sigma,u,q^2)&=-(1-\sigma)^2 M_{\Omega_Q}\left(\overline{\psi}^n(\omega,u)-\overline{\psi}^{\overline{n}}(\omega,u)\right).
    \end{align}

   In the above equations, the new representation of the amplitudes are defined by the following equation:
     \begin{gather}
     	\overline{\psi}_i (\omega,u)=\int_0^\omega d\tau \tau \tilde{\psi}_i (\tau,u).  
         \end{gather}

The relations among the form factors are given by $g_1(q^2)=g_2(q^2)=0$ and $f_2(q^2)=f_1(q^2)$.
 
    With the form factors $f_i(q^2)$ and $g_i(q^2)$ obtained above, the decay width that we will calculate in the next section can be calculated and analyzed.
    
    \subsection{Semileptonic decay of $\Omega_Q$ to $\Xi$ baryon}
    
	With the analytical expressions for the $\Omega_Q\to\Xi$ transition form factors, we can now proceed to calculate the branching fractions of the semileptonic decays $\Omega_Q\to \Xi \ell \nu_\ell$. To determine the branching fractions for semileptonic decays of double strangeness heavy flavor baryons, it is necessary to know the $q^2$ dependence of the decay width, i.e., the differential decay width. For the semileptonic decays of double strangeness heavy flavor baryons, the differential decay width can be expressed as the sum of two polarized differential decay widths,
   \begin{gather}
    	\frac{d\Gamma}{dq^2}=\frac{d\Gamma_L}{dq^2}+\frac{d\Gamma_T}{dq^2}, \label{difwid}
    \end{gather}
    and the total decay width is
    \begin{gather}
    	\Gamma=\int_{m_\ell^2}^{(M_{\Omega_Q}-M_\Xi)^2}dq^2\frac{d\Gamma}{dq^2},
    \end{gather}
    where
    \begin{align}
         \frac{d\Gamma_L}{dq^2}&=\frac{G_F^2|V_{cd(ub)}|^2 q^2 p}{192\pi^3 M_{\Omega_Q}^2}(|H_{\frac{1}{2},0}|^2+|H_{-\frac{1}{2},0}|^2),  \\
    	 \frac{d\Gamma_T}{dq^2}&=\frac{G_F^2|V_{cd(ub)}|^2 q^2 p}{192\pi^3 M_{\Omega_Q}^2}(|H_{\frac{1}{2},1}|^2+|H_{-\frac{1}{2},-1}|^2).
    \end{align}

    In the above equations, $p=\sqrt{Q_+Q_-}/2M_{\Omega_Q}, Q_\pm=(M_{\Omega_Q}\pm M_\Xi)^2-q^2$, and $m_\ell$ is the mass of lepton. The related expressions of helicity amplitudes connected with form factors are given by
    \begin{align}
    H_{\frac{1}{2},0}^V=&-i\frac{\sqrt{Q_-}}{\sqrt{q^2}}[(M_{\Omega_Q}+M_\Xi)f_1-\frac{q^2}{M_{\Omega_Q}}f_2], \\
    H_{\frac{1}{2},1}^V=&i\sqrt{2Q_-}[-f_1+\frac{M_{\Omega_Q}+M_\Xi}{M_{\Omega_Q}}f_2],  \\
    H_{\frac{1}{2},0}^A=&-i\frac{\sqrt{Q_+}}{\sqrt{q^2}}[(M_{\Omega_Q}-M_\Xi)g_1+\frac{q^2}{M_{\Omega_Q}}g_2], \\
    H_{\frac{1}{2},1}^A=&i\sqrt{2Q_+}(-g_1-\frac{M_{\Omega_Q}-M_\Xi}{M_{\Omega_Q}}g_2).
    \end{align}
    The negative helicity amplitudes can be given by the positive helicity amplitudes as
    \begin{gather}
    	H_{-\lambda,-\lambda_W}^V=H_{\lambda,\lambda_W}^V,   \quad     
    	H_{-\lambda,-\lambda_W}^A=-H_{\lambda,\lambda_W}^A.
    \end{gather}
 $\lambda$ and $\lambda_W$ are the polarizations of the final $\Xi$ baryon and W Boson, respectively. $H^V$ and $H^A$ correspond to the vector and axial-vector form of helicity amplitudes.
 
    For the $V$-$A$ current, the total helicity amplitude is given by the difference between the vector and axial-vector helicity amplitudes,
    \begin{gather}
    	H_{\lambda,\lambda_W}=H_{\lambda,\lambda_W}^V-H_{\lambda,\lambda_W}^A.
    \end{gather}
	
	\section{Numerical Analysis.} \label{sec:III}

For the numerical analysis, we adopt the fundamental mass parameters and lifetimes of quarks, baryons, and mesons from the Particle Data Group~\cite{ParticleDataGroup:2024cfk},
     \begin{gather}
     	m_u=2.16~{\rm{MeV}}, \quad m_c=1.27~{\rm{GeV}},\quad m_b=4.18^{+0.04}_{-0.03}~{\rm{GeV}}, \nonumber \\
       M_{\Omega_b^-}=6.0461~{\rm{GeV}}, \quad M_{\Omega_c^0}=2.6952~{\rm{GeV}}, \nonumber \\
       M_{\Xi^0}=1.31486~{\rm{GeV}},  \quad M_{\Xi^-}=1.32171~{\rm{GeV}} , \nonumber \\
      M_{B_u}=5.2794~{\rm{GeV}}, \quad   M_{B_d}=1.86966~{\rm{GeV}}, \nonumber \\
     \tau_{\Omega_b^-}=1.65^{+0.18}_{-0.16}\times10^{-12}~\rm{s}, \nonumber \\ 
     \tau_{\Omega_c^0}=(2.73\pm0.12)\times 10^{-13}~{\rm{s}}, \nonumber
     \end{gather}
   and the CKM matrix elements 
   $$|V_{cd}|=0.221\pm0.004$$ and $$|V_{ub}|=(3.82\pm0.20)\times 10^{-3}.$$

    In the QCD LCSRs, the threshold parameter $s_0$ introduced in matching the hadronic and QCD representations of correlation functions through dispersion relations can be constrained by the mass spectrum calculated using QCD sum rules. The threshold should effectively separate the ground-state baryon from excited states while ensuring that contributions to the form factors from the continuum above $s_0$ remain minimal. 

For the Borel parameter $M_B$ introduced via the Borel transformation, we require that it yields stable form factor values at fixed momentum-transfer squared $q^2$. These requirements translate to the following: (i) the ground-state contribution below $s_0$ should exceed 70\% in the sum rules, the contribution from the highest-twist term in baryon LCDAs should be less than 15\%, specifically for the $\Xi$ baryon the twist-6 LCDA contribution should be below 15\% for $\Omega_c\to\Xi$; (ii) the ground-state contribution below $s_0$ should exceed 90\% in the sum rules, the contribution from the highest-twist-6 term in $\Xi$ baryon LCDAs should be less than 30\% for $\Omega_b\to\Xi$.

\begin{figure}[htbp]
    	\begin{center}
            \includegraphics[width=0.4\textwidth]{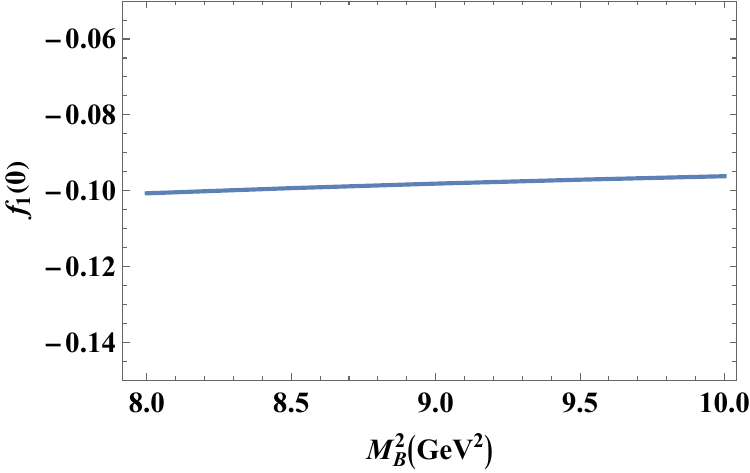}
                \qquad	
 		\includegraphics[width=0.4\textwidth]{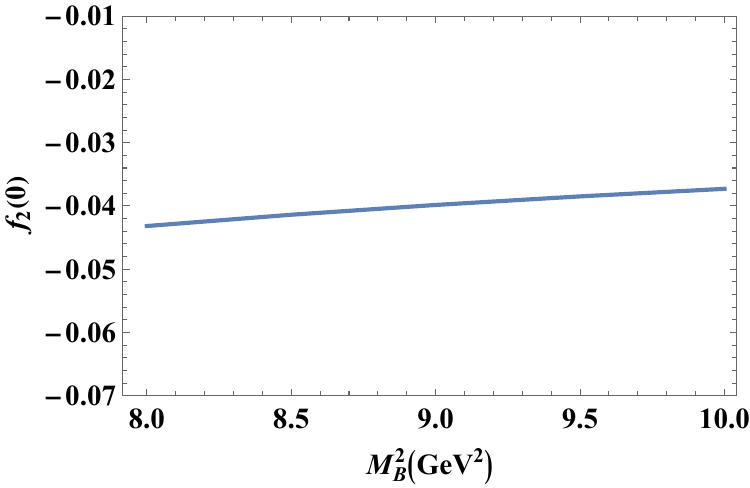}
    	\end{center}
         \caption{The Borel parameter dependence of the $\Omega_c^0 \to \Xi^-$ transition form factors at $q^2=0~\rm{GeV^2}$  within the $\Xi$ baryon LCDAs.} \label{boreldep1} 
    \end{figure}

\begin{figure}[htbp]
    	\begin{center}
            \includegraphics[width=0.4\textwidth]{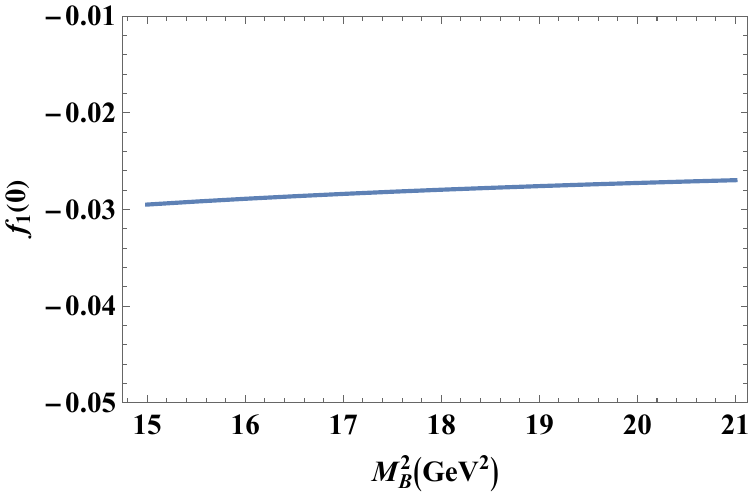}
                \qquad	
 		\includegraphics[width=0.4\textwidth]{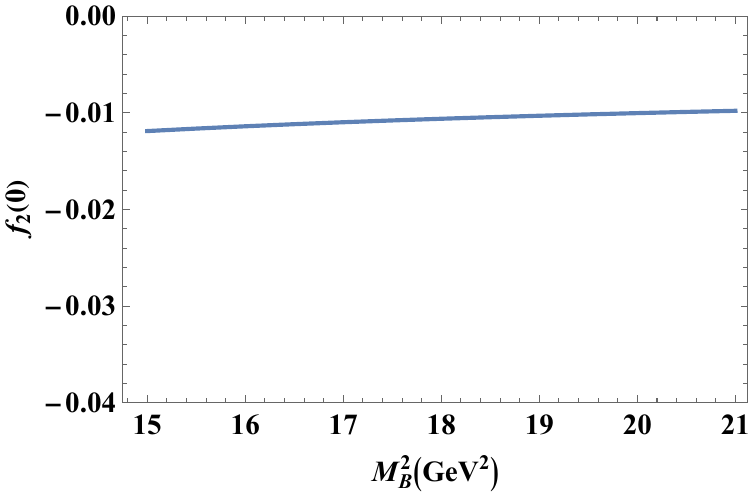}
    	\end{center}
         \caption{The Borel parameter dependence of the $\Omega_b^- \to \Xi^0$ transition form factors at $q^2=0~\rm{GeV^2}$  within the $\Xi$ baryon LCDAs.} \label{boreldep2} 
    \end{figure}

Accordingly, we choose $s_0 =( M_{\Omega_Q} + \Delta)^2$ with $\Delta = (0.5\pm0.1)$ GeV,  $M_B^2 = (10 \pm 1)$ GeV$^2$ for $\Omega_c^0\to\Xi^-$, and $M_B^2=(18\pm3)\rm{GeV^2}$ for $\Omega_b^-\to\Xi^0$ when using $\Xi$ baryon LCDAs. The Borel parameter dependence of the form factors at $q^2=0~\rm{GeV^2}$ are shown in Fig.~\ref{boreldep1} ($\Omega_c \to \Xi$) and Fig.~\ref{boreldep2} ($\Omega_b \to \Xi$). The baryon decay constants appearing in the sum rules are taken from previous QCD sum rule calculations: $f_{\Omega_b^-} = 0.134$ GeV$^3$, $f_{\Omega_c^0} = 0.093$ GeV$^3$, and $f_\Xi = 0.0099$ GeV$^3$~\cite{Wang:2009cr,Liu:2009uc}.

\begin{figure}[htbp]
    	\begin{center}
            \includegraphics[width=0.4\textwidth]{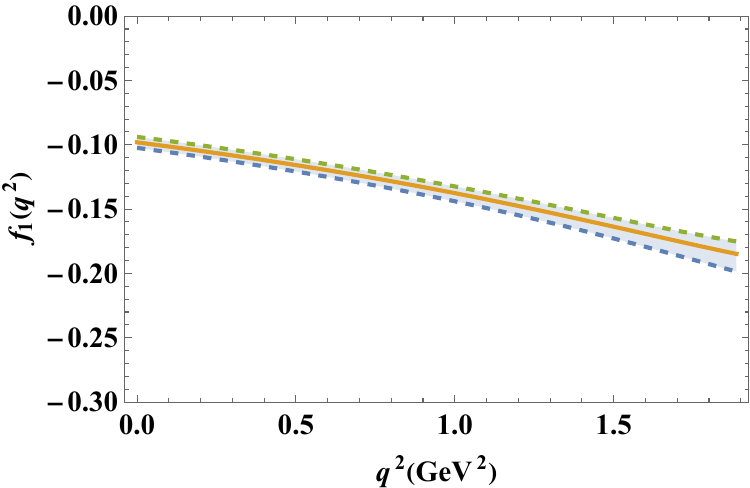}
                \qquad	
 		\includegraphics[width=0.4\textwidth]{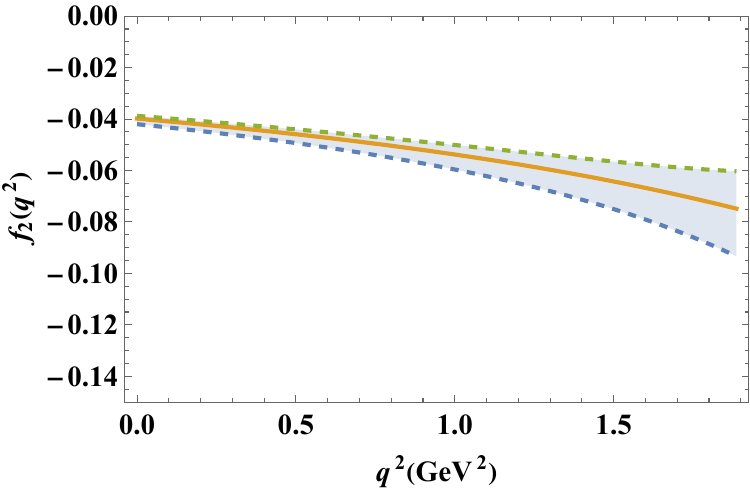}
    	\end{center}
         \caption{Form factors  $f_i (i=1, 2)$ of $\Omega_c^0$ transition to $\Xi^-$ with $q^2$ dependence within $\Xi$ baryon LCDAs.} \label{ffc} 
    \end{figure}

\begin{figure}[htbp]
    	\begin{center}
            \includegraphics[width=0.4\textwidth]{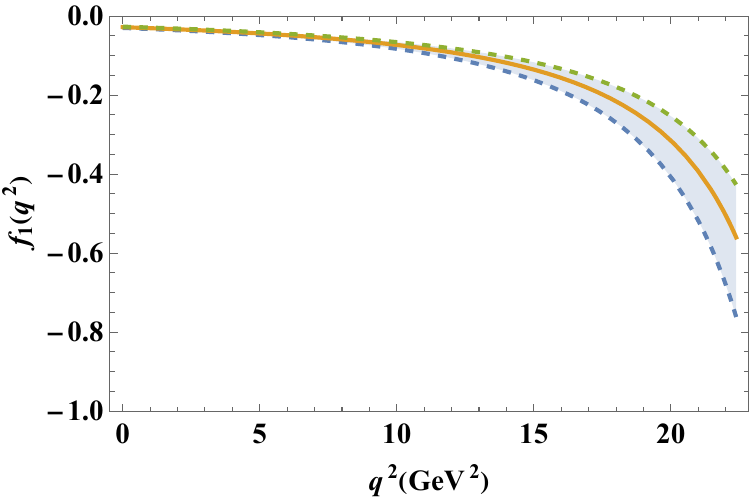}
                \qquad	
 		\includegraphics[width=0.4\textwidth]{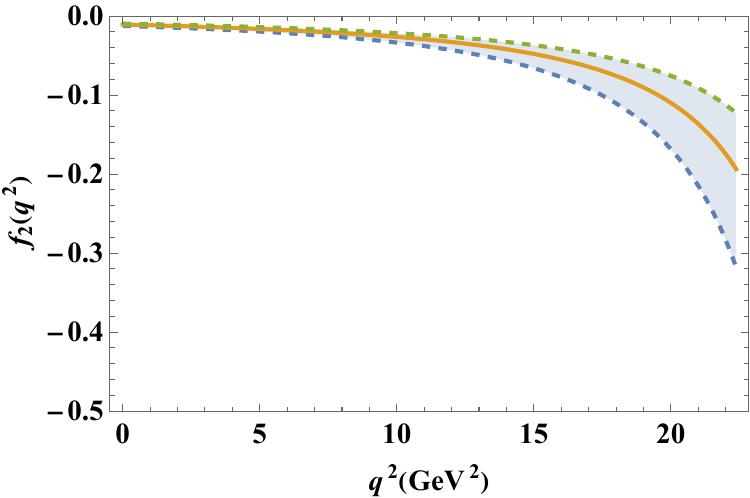}
    	\end{center}
         \caption{Form factors $f_i (i=1,2)$ of $\Omega_b^-$ transition to $\Xi^0$  with $q^2$ dependence within $\Xi$ baryon LCDAs.}  \label{ffb}
    \end{figure}

The form factors calculated using $\Xi$ baryon LCDAs can be plotted as functions of $q^2$ with these parameter choices. Note that, while the physical range for $\Omega_Q \to \Xi$ semileptonic decays is $m_\ell^2 < q^2 < (M_{\Omega_Q} - M_\Xi)^2$, the LCSR results are only reliable near $q^2 \approx 0$ GeV$^2$. To extend the form factors to the entire physical region, we employ the $z$-series expansion formula
 \begin{align}
 f_i(q^2)=\frac{f_i(0)}{1-q^2/M_{B_u/D_d}^2}\left[1+a\left(z(q^2)-z(0)\right)
 +b\left(z^2(q^2)-z^2(0)\right)\right],
     \end{align}
   where
\begin{align}
z(q^2)=\frac{\sqrt{M_{B_u/D_d}^2-q^2}-\sqrt{M_{B_u/D_d}^2-(M_{\Omega_Q}-M_{\Xi})^2}}{\sqrt{M_{B_u/D_d}^2-q^2}+\sqrt{M_{B_u/D_d}^2-(M_{\Omega_Q}-M_{\Xi})^2}}.
\end{align}
Here $a$ and $b$ are fitting parameters, and $M_{D_d}$, $M_{B_u}$ represent the masses of $D_d$ and $B_u$ mesons. In our analysis,
\begin{itemize}
\item[(i)] $\Omega_c^0 \to \Xi^-$: LCSR valid in $0 < q^2 < 1$ GeV$^2$;
\item[(ii)] $\Omega_b^- \to \Xi^0$: LCSR valid in $0 < q^2 < 10$ GeV$^2$.
\end{itemize}
The corresponding form factor extrapolations are shown in Figs.~\ref{ffc} and~\ref{ffb}, while their values at $q^2=0~\rm{GeV^2}$ and fitting parameters are listed in Tables~\ref{fffitc} and~\ref{fffitb}.  

    \begin{table}[htbp]
     	\centering
     	\caption{The fitting formula parameters $\it a$ and $\it b$ of $\Omega_c^0 \to \Xi^-$ form factors within $\Xi$ baryon LCDAs.}  
     	\begin{tabular}{cccc}\hline
     		$f_i(q^2)$&           $f_i(0)$                        &                    $\it a$                  &          $\it b$                            \\ \hline
     		$f_1(q^2)$&$-0.098^{+0.004}_{-0.004}$&$1.934^{+0.240}_{-0.323}$&$-6.473^{+0.985}_{-1.031}$           \\ 
     		$f_2(q^2)$&$-0.040^{+0.001}_{-0.002}$&$1.179^{+1.255}_{-1.229}$ &$-2.471^{+2.092}_{-2.493}$          \\ \hline
     	\end{tabular} \label{fffitc}
     \end{table}
     
The central values in Table~\ref{fffitc} correspond to the parameters $s_0=(M_\Xi+0.5)^2~\rm{GeV^2}$ and $M_B^2=9~\rm{GeV^2}$. The upper bounds of $f_i(0)$ and $a$, along with the lower bounds of $b$, are obtained with $s_0=(M_\Xi+0.4)^2~\rm{GeV^2}$ and $M_B^2=10~\rm{GeV^2}$. Conversely, the lower bounds of $f_i(0)$ and $a$, together with the upper bounds of $b$, correspond to $s_0=(M_\Xi+0.6)^2~\rm{GeV^2}$ and $M_B^2=8~\rm{GeV^2}$.

  \begin{table}[htbp]
     	\centering
     	\caption{The fitting formula parameters $\it a$ and $\it b$ of $\Omega_b^- \to \Xi^0$ form factors within $\Xi$ baryon LCDAs.}  
     	\begin{tabular}{cccc}\hline
     		$f_i(q^2)$&           $f_i(0)$                        &                    $\it a$                  &          $\it b$                            \\ \hline
     		$f_1(q^2)$&$-0.028^{+0.001}_{-0.002}$&$-8.737^{+3.608}_{-5.123}$&$2.845^{+5.925}_{-4.042}$           \\ 
     		$f_2(q^2)$&$-0.011^{+0.002}_{-0.001}$&$-7.773^{+4.649}_{-6.918}$ &$2.710^{+7.991}_{-5.199}$          \\ \hline
     	\end{tabular} 
     	\label{fffitb}
     \end{table}

The central values in Table~\ref{fffitb} correspond to the parameters $s_0=(M_\Xi+0.5)^2~\rm{GeV^2}$, $M_B^2=18~\rm{GeV^2}$ and $m_b=4.18~\rm{GeV}$. The upper bounds of $f_i(0)$ and $a$, along with the lower bounds of $b$, are obtained with $s_0=(M_\Xi+0.6)^2~\rm{GeV^2}$, $M_B^2=21~\rm{GeV^2}$ and $m_b=4.22~\rm{GeV}$. Conversely, the lower bounds of $f_i(0)$ and $a$, together with the upper bounds of $b$, correspond to $s_0=(M_\Xi+0.4)^2~\rm{GeV^2}$, $M_B^2=15~\rm{GeV^2}$ and $m_b=4.15~\rm{GeV}$.

     \begin{figure}[htbp]
     	\begin{center}
    		\includegraphics[width=0.5\textwidth]{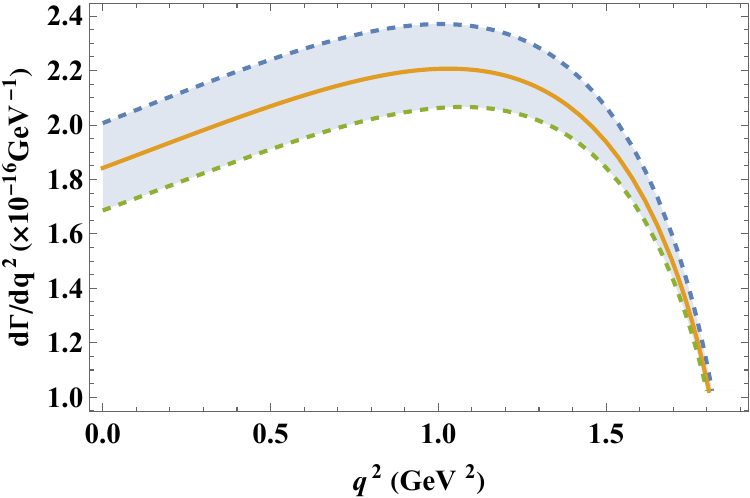}     		
     	\end{center}
     	\caption{The differential decay width of $\Omega_c^0\to\Xi^-\ell^+\bar{\nu}_\ell$ within $\Xi$ baryon LCDAs.}  \label{difwidc}
       \end{figure}    

\begin{table}[htbp]
\centering
\caption{Decay width and branching fractions of $\Omega_c^0$ transition to $\Xi^-$ semileptonic decay.} 
    \begin{tabular}{ccc} \hline
    Decay modes &$\Gamma (\times 10^{-16})$&$\mathcal{B}r (\times 10^{-4})$  \\ \hline
   $\Omega_c^0 \to \Xi^- \ell^+ \bar{\nu}_\ell$&$3.662^{+0.278}_{-0.240}$ &$1.519^{+0.115}_{-0.100}$ \\ 
\hline
\end{tabular} \label{tdifwidc}
\end{table} 

 \begin{figure}[htbp]
     	\begin{center}
    		\includegraphics[width=0.5\textwidth]{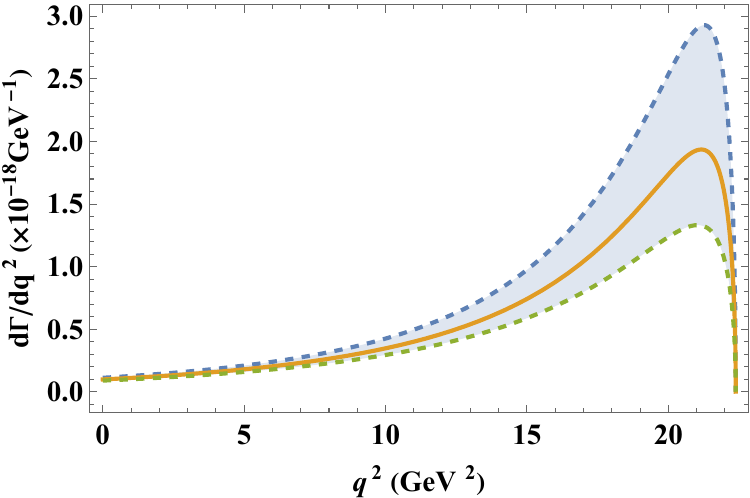} 
     	\end{center}
     	\caption{The differential decay width of $\Omega_b^-\to\Xi^0\ell^-\nu_\ell$ within $\Xi$ baryon LCDAs.}  \label{difwidb}
     \end{figure}   

\begin{table}[htbp]
\centering
\caption{Decay width and branching fractions of $\Omega_b$ transition to $\Xi$ semileptonic decay.} 
\begin{tabular}{ccc} \hline
    Decay modes &$\Gamma (\times 10^{-17})$&$\mathcal{B}r (\times 10^{-5})$  \\ \hline
   $\Omega_b^- \to \Xi^0 \ell^- \nu_\ell$&$1.451^{+0.544}_{-0.346}$ &$3.638^{+1.364}_{-0.869}$ \\ 
\hline
\end{tabular} \label{tdifwidb}
\end{table}

Substituting the fitted form factors into the differential decay width formula Eq.~(\ref{difwid}), we obtain the $q^2$ distributions shown in Fig.~\ref{difwidc} for $\Omega_c^0\to\Xi^-\ell^+\bar{\nu}_\ell$ and Fig.~\ref{difwidb} for $\Omega_b^-\to\Xi^0\ell^-\nu_\ell$. Integrating over the physical range yields the partial decay widths for $\Omega_Q \to \Xi\ell\nu_\ell$ processes. Combining with $\Omega_Q$ lifetimes gives the decay width and absolute branching fractions in Tables~\ref{tdifwidc} and~\ref{tdifwidb}. 

	    \begin{table}[htbp]
    	\centering
    	\caption{The decay width ($\Gamma$) obtained by LCSRs within light baryon $\Xi$ LCDAs and compared with other approaches.}  
    	\begin{tabular}{cccc}\hline
    \multirow{2}*{Decay modes}                       &                                   \multicolumn{3}{c}{Decay width $\Gamma$ ($\rm{GeV}$) }             \\
    		                                                            &       This work                                                    &            Ref.~\cite{Pervin:2006ie}                         & Ref.~\cite{Zhao:2018zcb} \\ \hline
     $\Omega_c^0\to\Xi^- \ell^+ \bar{\nu}_\ell$  & $3.662^{+0.278}_{-0.240}\times 10^{-16}$ & $(0.34\sim0.65)\times 6.582\times10^{-15}$ &$2.08\times10^{-15}$ \\
    $\Omega_b^-\to\Xi^0 \ell^-{\nu}_\ell$&$1.451^{+0.544}_{-0.346}\times 10^{-17}$  & $(0.82\sim1.78)\times 2.514\times 10^{-18}$&$1.18\times10^{-17}$ \\ \hline
    	\end{tabular} \label{comp}
    \end{table}

Current experimental results for these decay channels remain unavailable, and theoretical calculations are scarce. Our results show good agreement with the light-front quark model predictions under SU(3) symmetry in Ref.~\cite{Zhao:2018zcb}. The comparison of our results with other's are listed in Table~\ref{comp}. The value between two times for $\Omega_b^-\to\Xi^0\ell^-\nu_\ell$ of the Ref.~\cite{Pervin:2006ie} column comes from CKM matrix element $|V_{ub}|$ and Planck constant $\hbar$;  the value between two times for $\Omega_c^0\to\Xi^-\ell^+\bar{\nu}_\ell$ comes from the Planck constant $\hbar$. 

When analyzing these decay channels using the LCDAs of $\Omega_Q$ baryons, we observe that the twist-4 $\Omega_Q$ LCDAs dominate the contributions. Notably, the obtained branching fraction for $\Omega_b^-\to\Xi^0\ell^- \nu_\ell$ and $\Omega_c^0\to\Xi^-\ell^+\bar{\nu}_\ell$ are nearly 3 orders of magnitude larger than the results derived from $\Xi$ baryon LCDAs. When requiring the contribution of the ground-state hadron to exceed $60\%$ for $\Omega_b$ to $\Xi$ decay, on the Borel region $2<M_B^2<3~\rm{GeV^2}$ and $A$ from $0$ to $1$, the form factor at $q^2 = 0~\rm{GeV}^2$ is $f_1(0)=0.715^{+0.178}_{-0.008}$, and the corresponding absolute branching fraction $\mathcal{B}r(\Omega_b^- \to \Xi^0 \ell^- \nu_\ell)=(1.648^{+0.399}_{-0.267})\%$. In the case of $\Omega_c \to \Xi$, on the Borel region $4<M_B^2<10~\rm{GeV^2}$, we analyzed the form factor in the range $0 < q^2 < 0.6~\rm{GeV}^2$, also considering variations in $A$ from $0$ to $1$. The resulting form factor at $q^2 = 0$ is $f_1(0)=2.747^{+0.361}_{-0.166}$, leading to an absolute branching fraction of $14.23^{+8.58}_{-3.50}\%$ for the semileptonic decay $\Omega_c^0 \to \Xi^- \ell^+ \bar{\nu}_\ell$.

 These discrepancies show consistency with the differences calculated between various theoretical approaches for analyzing  the corresponding $\Omega_Q\to\Omega$ semileptonic decay channel in literature \cite{Aliev:2022gxi}. However, the almost 3-order-of-magnitude enhancement relative to $\Xi$ LCDAs results and their marked inconsistency with other theoretical calculations cast substantial doubt on the reliability of these findings. The feasibility of extracting reliable predictions using $\Omega_Q$ LCDAs requires further investigation, particularly concerning the LCDAs of double strangeness heavy flavor baryons and the QCD LCSRs framework for these specific processes. These aspects will be addressed in our future work.

	\section{Conclusions and discussions} 
	
     In this work, we calculate the form factors of double strangeness heavy flavor baryon decays into $\Xi$ baryons and the corresponding absolute branching ratios of semileptonic decay processes using QCD LCSRs. When employing one version of LCSRs—specifically, analyzing the LCDAs of the final-state $\Xi$ baryon—our results agree with those obtained by other authors using the light-front quark model~\cite{Zhao:2018zcb}. However, when adopting an alternative approach within LCSRs, namely analyzing the same processes using the LCDAs of the initial double strangeness heavy flavor baryon, we find that the branching ratios of $\Omega_b^-\to\Xi^0\ell^-\nu_\ell$ and $\Omega_c^0 \to \Xi^- \ell^+ \bar{\nu}_\ell$ become almost 3 orders of magnitude larger than the previous result and inconsistent with the light-front quark model prediction.

Our analysis reveals that the LCDAs of heavy flavor baryons have been successfully applied in QCD LCSRs for calculating weak decay form factors and semileptonic branching ratios in cases involving either no strange quarks or a single strange quark \cite{Duan:2024lnw, Duan:2022uzm, Aliev:2023mdf}. However, when employing LCDAs of double strangeness heavy flavor baryons (containing two strange quarks), the resulting semileptonic branching ratios exhibit order-of-magnitude discrepancies compared with other theoretical approaches. This discrepancy may originate from the current theoretical limitations in LCDA development. While the $\Xi$ baryon LCDAs have been theoretically investigated up to twist 6, the LCDAs for double strangeness heavy flavor baryons have only been studied up to twist 4 at present.

These findings necessitate further systematic investigations of the LCDAs for double strangeness heavy flavor baryons in our subsequent work. In particular, comprehensive studies are required to examine the fundamental differences between LCDAs of double strangeness baryons and those containing either zero or one strange quark and quantify the contributions from higher-twist (twist 5 and beyond) LCDAs of heavy flavor baryons in QCD LCSRs calculations.

	\section*{Acknowledgments}
H.H. D is very greatful for the help from Professor Y.M. Wang.This work was supported by the Postdoctoral Fellowship Program of CPSF under Grant No. GZC20230738, National Natural Science Foundation of China (Grant No. 12275067), Science and Technology R$\&$D Program Joint Fund Project of Henan Province  (Grant No. 225200810030), and Science and Technology Innovation Leading Talent Support Program of Henan Province  (Grant No. 254000510039). 
	
       \bibliography{references}
	\end{document}